\pdfoutput=1

\documentclass[11pt]{article}

\usepackage[preprint]{acl}

\usepackage{times}
\usepackage{latexsym}
\usepackage{enumitem}
\usepackage{booktabs}
\usepackage[edges]{forest}
\usepackage{array}
\usepackage[T1]{fontenc}

\usepackage[utf8]{inputenc}
\usepackage{booktabs}
\usepackage{tabularx}
\usepackage{multirow}
\usepackage{rotating}

\usepackage{microtype}

\usepackage{inconsolata}

\usepackage{graphicx}
\usepackage{amsmath}

\usepackage{tikz}
\usepackage{xcolor}
\usepackage{multirow}
\usepackage{natbib}
\usepackage{mydef}
\usepackage{xspace}
\usepackage{pifont}

\newcommand{\proteinllms}{Protein LLMs\xspace}

\title{Protein Large Language Models: A Comprehensive Survey}

\author{%
    \begin{tabular}{c}
        Yijia Xiao\textsuperscript{\ding{171}} \thanks{Contact Email: \texttt{yijia.xiao@cs.ucla.edu}}, Wanjia Zhao\textsuperscript{\ding{168}}, Junkai Zhang\textsuperscript{\ding{171}}, 
        Yiqiao Jin\textsuperscript{\ding{170}}, Han Zhang\textsuperscript{\ding{171}}, \\
        Zhicheng Ren\textsuperscript{\ding{171}},  
        Renliang Sun\textsuperscript{\ding{171}}, Haixin Wang\textsuperscript{\ding{171}}, 
        Guancheng Wan\textsuperscript{\ding{171}}, Pan Lu\textsuperscript{\ding{168}}, \\
        Xiao Luo\textsuperscript{\ding{171}},  
        Yu Zhang\textsuperscript{\ding{169}}, James Zou\textsuperscript{\ding{168}}, Yizhou Sun\textsuperscript{\ding{171}}, Wei Wang\textsuperscript{\ding{171}} \\
    \end{tabular}\\[3.5ex]
    \textsuperscript{\ding{171}}UCLA, \quad 
    \textsuperscript{\ding{168}}Stanford, \quad 
    \textsuperscript{\ding{170}}Georgia Tech, \quad 
    \textsuperscript{\ding{169}}Texas A\&M  \\[1ex]
    \texttt{\url{https://github.com/Yijia-Xiao/Protein-LLM-Survey}}
}

\definecolor{myred}{RGB}{247,226,231}
\definecolor{myblue}{RGB}{216,226,234}
\definecolor{myyellow}{RGB}{252,238,221}
\definecolor{mypurple}{RGB}{233,229,241}
\definecolor{mygreen}{RGB}{204,231,207}

\begin{document}
\maketitle

\begin{abstract}
Protein-specific large language models (\proteinllms) are revolutionizing protein science by enabling more efficient protein structure prediction, function annotation, and design. While existing surveys focus on specific aspects or applications, this work provides the first comprehensive overview of \proteinllms, covering their architectures, training datasets, evaluation metrics, and diverse applications. Through a systematic analysis of over 100 articles, we propose a structured taxonomy of state-of-the-art \proteinllms, analyze how they leverage large-scale protein sequence data for improved accuracy, and explore their potential in advancing protein engineering and biomedical research.  Additionally, we discuss key challenges and future directions, positioning \proteinllms as essential tools for scientific discovery in protein science. Resources are maintained at \url{https://github.com/Yijia-Xiao/Protein-LLM-Survey}.
\end{abstract}

\section{Introduction}

``\textit{Proteins are the machinery of life, and understanding their language unlocks the secrets of biology.}''
\rightline{--- David Baker (Nobel Prize laureate 2024)}
\\

Proteins are essential biological molecules, driving functions such as catalyzing biochemical reactions, maintaining cell structure, and enabling cellular communication. Understanding their sequence-structure-function relationships is central to biological research. However, traditional experimental methods, including X-ray crystallography, NMR spectroscopy, and cryo-electron microscopy, are time-consuming and labor-intensive, posing bottlenecks for large-scale applications.

Recent advancements in language modeling have revolutionized computational biology, offering powerful tools for protein analysis. Protein large language models (\textbf{\proteinllms}) share several foundational similarities with LLMs: 1) \emph{Training objectives and learning paradigms}, both LLMs and \proteinllms are trained in a self-supervised manner on large-scale datasets using objectives such as masked language modeling~\cite{devlin2019bert},  auto-regressive modeling~\cite{luo2022biogpt}, or sentence permutation~\cite{lewis2019bart, yuan2022biobart}, learning to predict missing or next elements in sequences from the vocabulary. While LLMs predict missing words or phrases within textual data~\cite{reimers2019sentence, liu2019roberta, touvron2023llama}, \proteinllms predict amino acids or subsequences within protein sequences. 2) \emph{Pretraining data.} \proteinllms adopt a data-driven paradigm to learn directly from large-scale protein datasets~\cite{liu2024timemattersexaminetemporal, jones2024examiningimbalanceeffectsperformance}. The datasets for training \proteinllms consist of vast collections of protein sequences, analogous to the textual corpora used for LLMs. This eliminates the need for explicit feature engineering, allowing \proteinllms to learn intricate patterns, such as structural motifs, evolutionary relationships, and functional insights, similar to how LLMs capture semantic and syntactic structures in language.

This paradigm shift has led to the emergence of highly effective models that can predict protein folding, annotate biological functions, and even design novel proteins with desired characteristics. Beyond their predictive capabilities, \proteinllms also provide interactive interfaces that allow users to upload protein sequences or structural files (e.g., PDB format), pose questions, and interact with the model in a conversational manner~\cite{liu2024prott3,xiao2024proteingpt}, proving deeper insights into protein structure, function, and design.

We present the first dedicated survey of \proteinllms, analyzing their unique architectures, training methodologies, and practical applications in protein research. While previous studies have explored the applications of various computational methods for protein research~\cite{xinhui2024generative, wu2022survey} or discussed the role of language models in general scientific domains such as biomedicine~\cite{wang2023pre} and chemistry~\cite{liao2024words}, this survey focuses specifically on \proteinllms--a rapidly evolving area at the intersection of computational biology and NLP.  

\begin{figure*}[ht]
\centering
\begin{forest}
  for tree={
    forked edges,
    grow=east,
    reversed=true,
    anchor=base west,
    parent anchor=east,
    child anchor=west,
    base=middle,
    font=\scriptsize,
    rectangle,
    draw=black,
    edge=black!50, 
    rounded corners,
    align=center,
    minimum width=1em,
    s sep=5pt,
    inner xsep=2.5pt,
    inner ysep=1pt
  },
  [Protein Large Language Models,rotate=90,anchor=north,edge=black!50,fill=myblue,draw=black
    [Problem,edge=black!50, fill=mypurple, minimum height=1.2em
      [Protein Understanding \& Prediction,text width=9.6em,fill=mypurple
        [\textbf{Pretraining Dataset:} UniRef Clusters~\cite{suzek2015uniref}{,} Pfam~\cite{finn2006pfam}{,} \\UniProtKB~\cite{boutet2016uniprotkb,m1999edittotrembl},text width=23em, fill=mypurple]
       [\textbf{Benchmark Dataset:} CASP~\cite{kryshtafovych2019critical}{,} TAPE\citep{rao2019evaluating}{,} \\ProteinLMBench~\cite{shen2024fine}{,} PEER~\cite{xu2022peer},text width=23em, fill=mypurple]
      ]
      [Protein Engineering \& \\Generation  \& Translation ,text width=6.8em,fill=mypurple
        [\textbf{Pretraining Dataset:} AlphaFoldDB~\cite{tunyasuvunakool2021highly},text width=25.8em, fill=mypurple]
       [\textbf{Benchmark Dataset:} ProteinGym~\cite{notin2024proteingym}{,} ProteinLMBench~\cite{shen2024fine},text width=25.8em, fill=mypurple]]
      ]
    [Method, edge=black!50, fill=myred, minimum height=1.2em
      [LLM Methods for \\ProteinUnderstanding \\and Prediction, edge=black!50, fill=myred
        [Protein Sequence Models, text width=6.6em, fill=myred
          [ UniRep~\citep{alley2019unified}{,} SSA \citep{bepler2019learning}{,}\\ MuPIPR \citep{zhou2020mutation}{,} ProtTrans\citep{elnaggar2021prottrans}{,}\\ AminoBERT\citep{chowdhury2022single}{,} Provis \citep{vig2020bertology}{,}  \\xTrimoPGLM~\citep{chen2024xtrimopglm}{,} CSCS~\citep{hie2021learning}{,}\\TCR-BERT \citep{wu2024tcr}{,} PeptideBert \citep{guntuboina2023peptidebert}{,}\\
          ESM-1b \citep{rives2021biological}{,} ESM-1v \citep{meier2021language}{,} \\
          AlphaFold2 \citep{jumper2021highly}{,} ESM-IF \citep{hsu2022learning}{,}\\
          ESM-2 \citep{lin2023evolutionary}{,} ProtTrans \citep{elnaggar2021prottrans}{,} \\
          ProteinLM \citep{xiao2021modeling}{,} 
 ProteinBERT~\citep{brandes2022proteinbert}{,}\\
          MSA Transformer \citep{rao2021msa}{,} Tranception \citep{notin2022tranception}{,} \\vcMSA~\citep{mcwhite2023leveraging}{,} Poet~\citep{truong2023poet} , text width=18.9em, fill=myred]
        ]
        [Structure-Integrated \&\\ Knowledge-Enhanced \\Models, text width=5.7em, fill=myred
          [OntoProtein \citep{zhang2022ontoprotein}{,} ESM-GearNet \citep{zhang2023systematic}{,}\\
          SaProt \citep{su2023saprot}{,} ProteinBERT \citep{brandes2022proteinbert}{,}\\ProteinCLIP \citep{wu2024proteinclip} , text width=19.8em, fill=myred]
        ]
        [Protein Description \&\\ Annotation Models, text width=5.6em, fill=myred
          [ProTranslator~\citep{xu2022protranslator}{,} BioTranslator~\citep{xu2023multilingual}{,}\\ProtST~\cite{xu2023protst}{,} Prot2Text~\citep{abdine2024prot2text}{,} \\BioT5~\citep{pei2023biot5}{,} ProtChatGPT~\citep{wang2024protchatgpt}{,}\\ProteinChat~\citep{guo2023proteinchat}{,} ProteinGPT~\cite{xiao2024proteingpt},text width=19.8em, fill=myred]
        ]
      ]
      [LLM Methods for \\Protein Engineering\\\& Generation \& Translation, edge=black!50, fill=myred
        [Protein Engineering Models, text width=7.4em, fill=myred
        [
        ProteinDT~\citep{liu2023text}{,} PLMeAE~\cite{plmeae}{,} \\ Toursynbio~\cite{shen2024toursynbio}, text width=16.3em, fill=myred
        ]
        ]
        [Protein Generation Models, text width=7.2em, fill=myred
          [ProGen \citep{madani2023large}{,} ProtGPT2 \citep{ferruz2022protgpt2}{,}\\
          ProGen2 \citep{nijkamp2023progen2}{,} ProLLaMA~\citep{lv2024prollama}{,}\\Ankh \citep{elnaggar2023ankhoptimizedproteinlanguage}{,}
          PAAG \citep{yuan2024annotation} {,} \\ Pinal \citep{dai2024toward} {,} IgLM \citep{shuai2023iglm}{,}\\
          PALM-H3 \citep{he2024novo}{,}  LM-D~\citep{10.5555/3618408.3620189}, text width=16.6em, fill=myred]
        ]
        [Protein Translation Models, text width=7em, fill=myred
          [ProstT5 \citep{heinzinger2023bilingual}{,} Fold2Seq \citep{cao2021fold2seq} {,} \\
          ProtAgents \citep{ghafarollahi2024protagents},
           text width=16.8em, fill=myred]
        ]
      ]
      [Traditional experimental methods, edge=black!50, fill=myred
        [X-ray \citep{jones1986using}{,} NMR \citep{shukla2023biomolecular}{,} Cryo-EM \citep{lyumkis2019challenges}{,}\\
        CryoDRGN~\citep{zhong2021cryodrgn}{,} CryoGAN~\citep{gupta2021cryogan}{,}\\
        CryoSTAR~\citep{li2024cryostar}{,} E2gmm~\citep{chen2021deep}, text width=24.1em, fill=myred]
      ]
    ]
    [Evaluation Metrics,edge=black!50, fill=mygreen, minimum height=1.2em
        [Structure Prediction Metrics, text width=7.6em,fill=mygreen
        [RMSD~\cite{li2013difficulty}{,} GDT-TS~\cite{zemla2003lga}{,} TM-Score~\cite{zhang2004scoring}{,}\\lDDT~\cite{mariani2013lddt}{,} pLDDT ~\cite{guo2022alphafold2}, text width=22.3em, fill=mygreen]
        ]
        [Function Prediction Metrics, text width=7.6em,fill=mygreen
        [Accuracy{,} Precision{,} Recall{,} F1-score{,} AUC{,} BLEU \cite{papineni2002bleu}{,} \\ROUGE-L \cite{lin2004rouge}{,} METEOR \cite{banerjee2005meteor}{,} \\Subcellular Localization~\cite{briesemeister2010yloc}{,} Stability~\cite{cheng2006prediction}{,}\\Homology Detection~\cite{altschul1990basic,hamamsy2024protein}{,} \\ Solubility~\cite{hebditch2017protein}{,} Mutation Effect Prediction~\cite{mansoor2022accurate},text width=22.3em, fill=mygreen]
	]
        [Sequence Generation Metrics, text width=7.6em,fill=mygreen
	    [Perplexity~\cite{hesslow2022rita}{,} Diversity~\cite{bywater2015prediction,mcgee2021generative}{,}\\Novelty~\cite{truong2023poet}{,} Fr\'echet Protein Distance~\citep{jiang2008protein}{,} \\Foldability~\cite{baek2021accurate,magliery2015protein}{,} Recovery~\cite{watson2023novo}, text width=22.3em, fill=mygreen]
            ]
      ]
  ]
\end{forest}
\caption{Taxonomy of Protein Large Language Models.}
\label{fig:taxonomy}
\end{figure*}
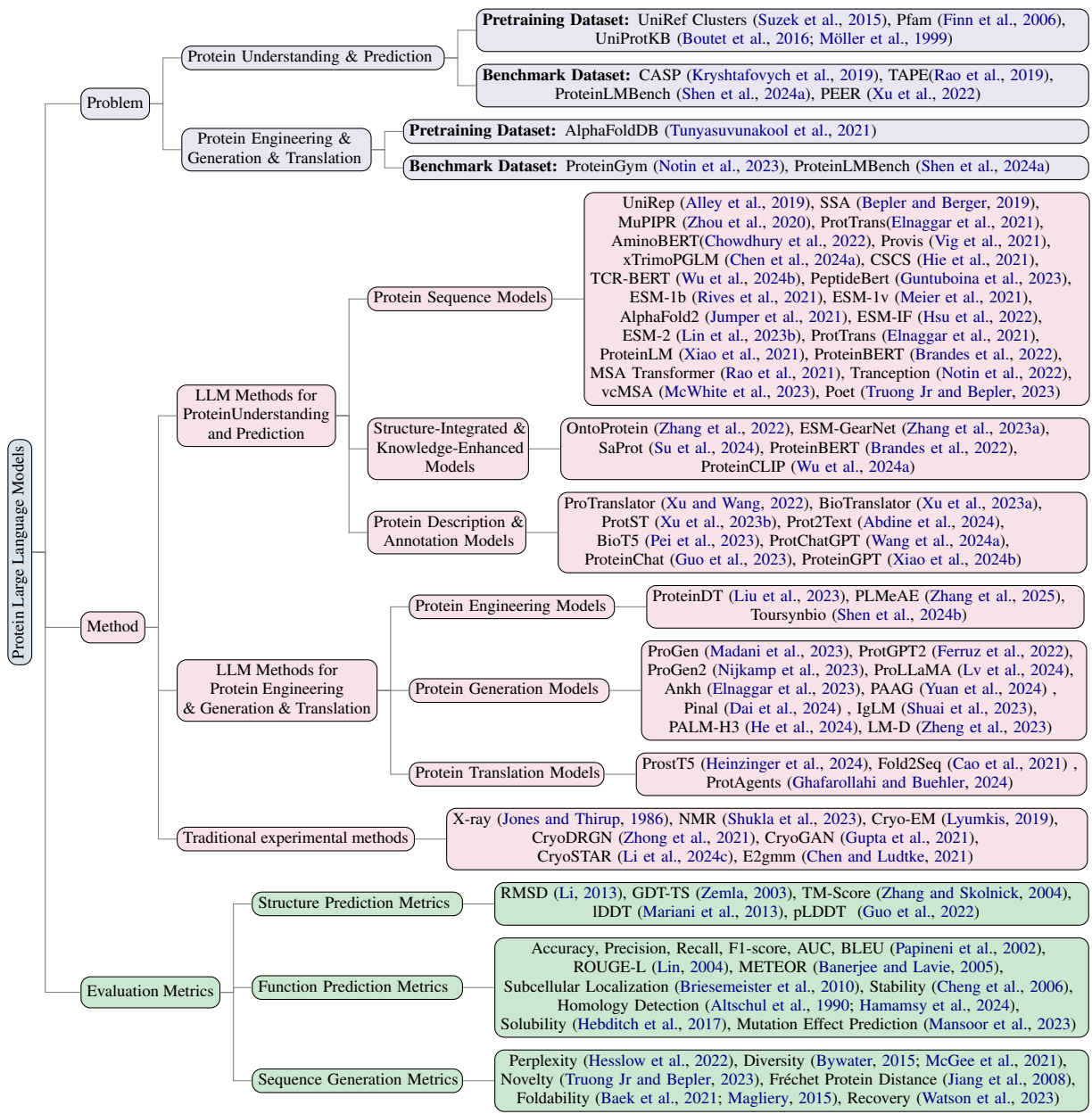

The key contributions are as follows:

\begin{itemize}[leftmargin=1em, noitemsep, topsep=0pt]
    \item \textbf{Architectural Overview.} A structured taxonomy of state-of-the-art \proteinllms (Figure \ref{fig:taxonomy}) detailing their unique architectures for protein understanding (\S\ref{sec:llm_understanding}) and generation (\S\ref{sec:llm_generation}), highlighting how these models surpass traditional experimental methods in both efficiency and accuracy (Appendix \S\ref{sec:experiment}). 
    \item \textbf{Data Insights.} A comprehensive summary of datasets for pretraining, fine-tuning, and benchmarking \proteinllms, providing critical insights into data curation strategies and their impact on model performance (\S\ref{sec:dataset}).
    \item \textbf{Evaluation Protocols.} A thorough discussion of methodologies for assessing the performance and impact of \proteinllms, including comprehensive new benchmarking strategies (\S\ref{sec:eval} and Appendix \S\ref{sec:app_eval}).
    \item \textbf{Applications.} A detailed exploration of practical applications in protein prediction, annotation, and design, remarkably highlighting recent innovative advancements and showcasing the transformative potential of \proteinllms in advancing biomedical research.
\end{itemize}

\begin{figure*}[htbp]
    \centering
    \includegraphics[width=\textwidth]{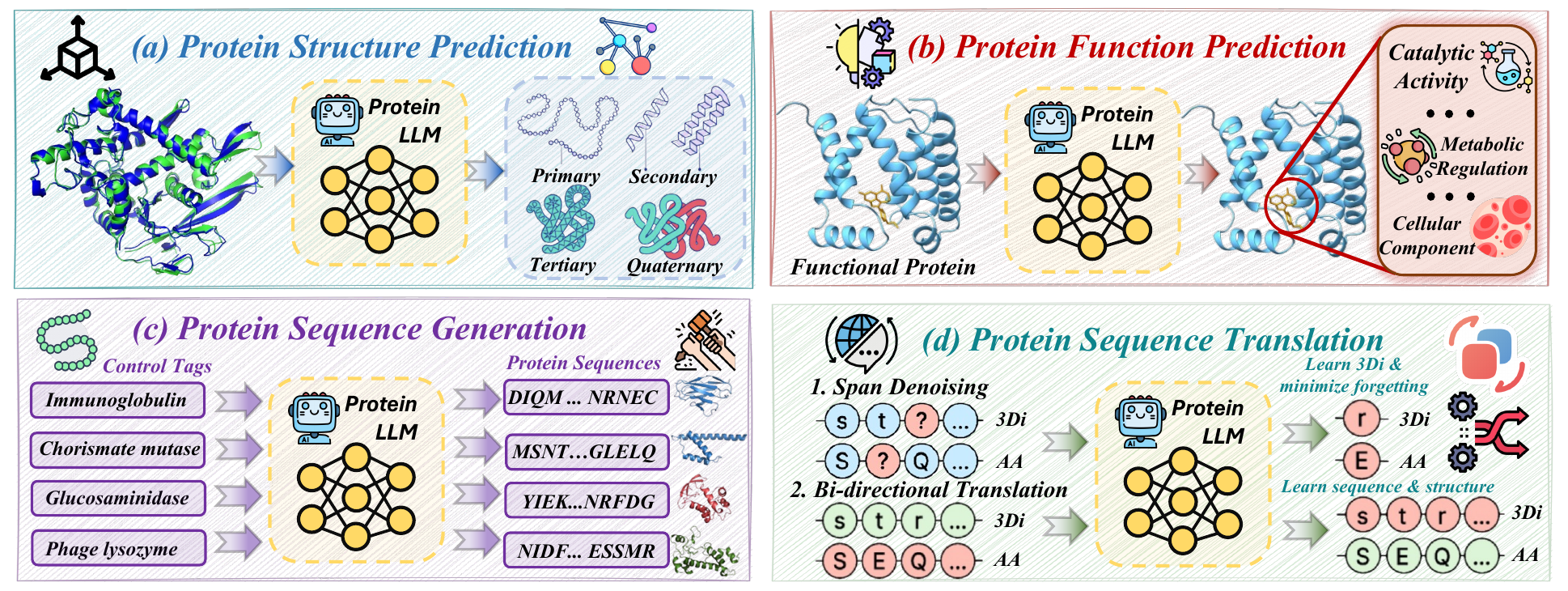}
    \caption{An Overview of Tasks in Protein Large Language Models.}
    \label{fig:methods}
\end{figure*}

\section{LLM Methods for Protein Understanding and Prediction}

\label{sec:llm_understanding}

\subsection{Problem Definition}

A protein, composed of amino acids (residues), can be represented as a sequence \( [x_1, \dots, x_L] \) in the residue token space \( \mathcal{P} \), where \(L\) denotes its length. According to Anfinsen's dogma, a protein’s primary sequence determines its structure and function. General problems in protein understanding and prediction are as follows:

\noindent \textit{I. Sequence-to-Property Prediction:} \( f_\theta: \mathcal{P} \rightarrow \mathcal{R}^+ \) mapping sequences to numerical properties, such as stability or fluorescence intensity.

\noindent \textit{II. Sequence-to-Label Prediction:} \( f_\theta: \mathcal{P} \rightarrow \mathcal{L} \) mapping sequences to categorical labels, including secondary structure types, contact maps, or functional annotations.

\noindent \textit{III. Sequence-to-Structure Prediction} \( f_\theta: \mathcal{P} \rightarrow \mathcal{S} \) mapping sequences to the 3D folding structures (i.e. tertiary structures). 

\noindent \textit{IV. Sequence-to-Text Understanding:} \( f_\theta: \mathcal{P} \rightarrow \mathcal{T} \), where \( \mathcal{T} \) represents generated textual descriptions of protein sequences.

\subsection{Protein Sequence Models}

\noindent \textbf{Individual Protein Sequences Models.}
Protein language models process amino acid sequences into meaningful representations for downstream tasks including structure and function prediction. Like NLP models, they are usually first pretrained on large sequence datasets with masked language modeling (MLM) objective; and then the protein sequences' embeddings are adapted for downstream tasks.
Initially, researchers leveraged long short-term memory (LSTM) architectures to learn representation of proteins \citep{alley2019unified,bepler2019learning,zhou2020mutation}. 
Following the breakthrough of transformer architectures \citep{vaswani2017attention} in NLP, transformer-based protein language models emerged as the new paradigm. Large-scale transformer models, scaling up to billions of parameters and trained on millions of protein sequences, have demonstrated remarkable effectiveness for protein understanding and prediction tasks \citep{rao2019evaluating,elnaggar2021prottrans, xiao2021modeling,hu2022exploring}, and 3D structure folding \citep{chowdhury2022single,fang2022helixfold,chen2024xtrimopglm}.
The interpretability of these \proteinllms has also been explored, with \citep{vig2020bertology} analyzing learned representations through the lens of attention.
Beyond general-purpose protein language models, several works have focused on domain-specific applications. For instance, \citet{hie2021learning} applied BiLSTM to model viral escape patterns; TCR-BERT \citep{wu2024tcr} specialized in T-cell receptor (TCR) analysis for improved TCR-antigen binding prediction; PeptideBERT \citep{guntuboina2023peptidebert} focused on predicting key properties of peptides; \citet{kroll2023turnover,yu2023enzyme} adapted ESM-1b for enzymatic function prediction.

\noindent \textbf{Multiple Sequence Alignments (MSA) Models.} MSA aligns homologous proteins within sequence space by mapping their residues to the coordinate framework of a designated seed sequence. MSA reveals evolutionary relationships between proteins and thus serves as a cornerstone of computational biology, particularly for mutation effects prediction \citep{ram2022few,hawkins2021msa}. The MSA Transformer \citep{rao2021msa} processed MSAs instead of single sequences. It used a modified axial attention mechanism \citep{ho2019axial,child2019generating} to model both intra- and inter-sequence relationships. In contrast, Tranception \citep{notin2022tranception}, was trained on individual non-aligned sequences but could leverage aligned sequences during inference. It extracted patterns from contiguous protein subsequences and improves fitness prediction by integrating MSAs retrieved at inference time. In specific subdomains, \citet{lin2023deep} developed a transfer learning framework that utilized ESM-MSA-1b for transmembrane protein complexes. Additionally, vcMSA~\citep{mcwhite2023leveraging} and Poet~\citep{truong2023poet} leveraged protein LLMs to identify MSAs or homologous sequences.

\noindent \textbf{Evolutionary Scale Modeling (ESM) Series.} ESM is a family of transformer models for protein modeling. 
ESM-1b~\cite{rives2021biological}, the first model in the series with up to 669.2 million parameters, was trained on 250 million protein sequences using a masked language modeling (MLM) objective and contains up to 669.2 million parameters. 
Building on this,  ESM-1v~\citep{meier2021language} focused on predicting the effects of mutations in zero-shot setting, while incorporating the MSA Transformer \citep{rao2021msa} for few-shot mutation prediction. 
Thanks to the success of AlphaFold2 \citep{jumper2021highly}, ESM-IF \citep{hsu2022learning} utilized predicted structures to train large models combining Geometric Vector Perceptron \citep{jing2020learning} with GNN or transformer on the inverse folding task that predicts protein strings from the 3D structures. The new general-purpose language protein model ESM-2 \citep{lin2023evolutionary} further scaled up the model size to 15 billion parameters and incorporated a folding head to create an end-to-end single-sequence structure prediction model ESMFold. The latest model ESM-3~\citep{hayes2025simulating} is a multimodal generative model with 98 billion parameters that could reason over protein sequences, structures, and functions. Using a chain-of-thought approach, it successfully designed a novel fluorescent protein far from any known fluorescent proteins.
\begin{figure*}[htbp]
    \centering
    \includegraphics[width=\textwidth]{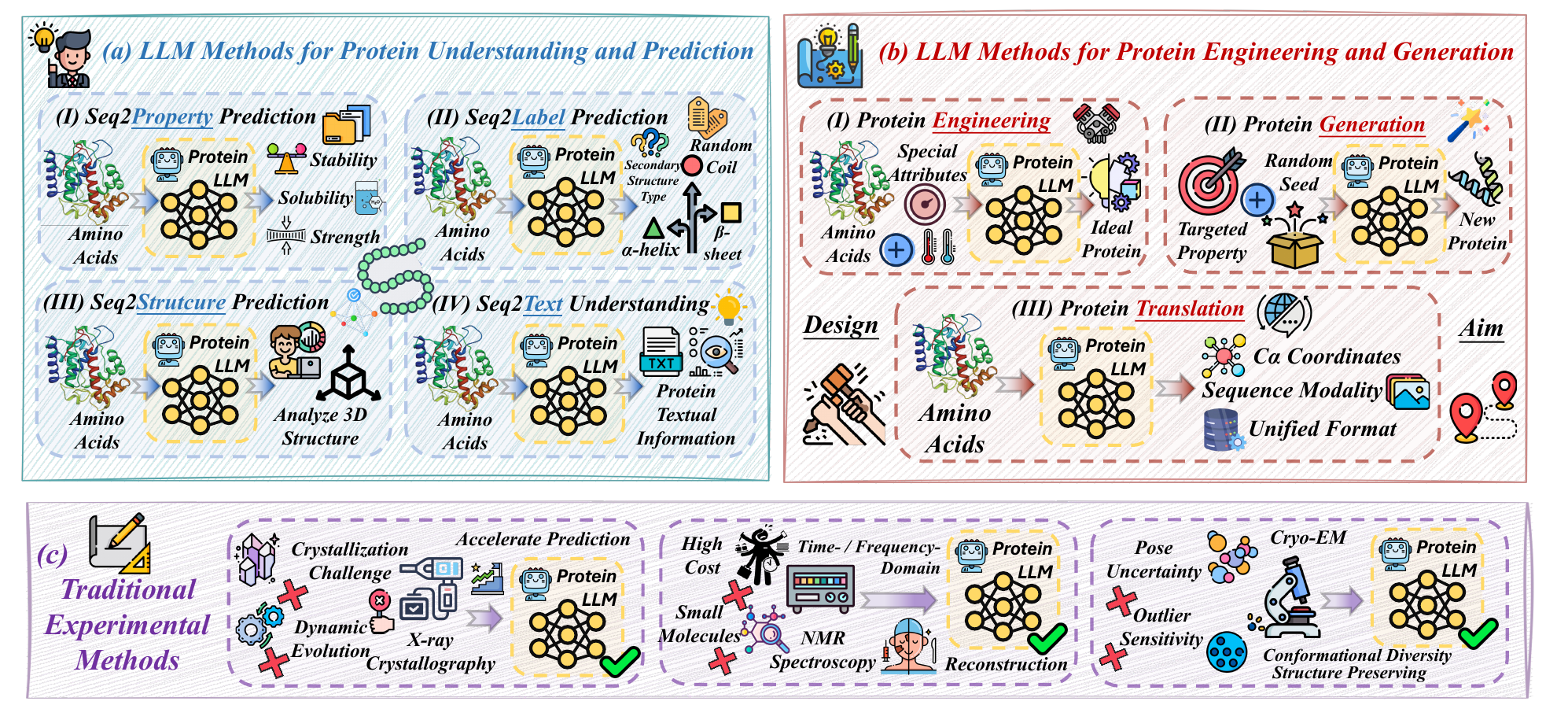}
    \caption{An Overview of Methods of Protein Large Language Models.}
    \label{fig:methods}
\end{figure*}

\subsection{Structure-Integrated and Knowledge-Enhanced Models}

Beyond residue sequences, many models integrate additional information, such as structure data or external knowledge, to enhance protein understanding and prediction ability.

\noindent \textbf{Structure-Integrated Models}: Structural information plays an important role in protein understanding, as a protein's functions are determined by its structures. Therefore, many works have incorporated structural information to enhance protein modeling ability. 
Some works utilized structure information as additional inputs \citep{chen2024endowing,tan2024simple}. For instance, \citet{zhang2023systematic} fused global structure information captured by structure encoder (GVP, GearNet \cite{zhang2022protein}, or CDConv \citep{fan2022continuous}) into representations of ESM-2; SaProt \citep{su2023saprot} incorporated local structural information for each amino acid, derived from Foldseek \citep{van2022foldseek}, to generate structure-aware tokens.
Alternatively, other works injected the structure information only in the training stage by either additional training tasks \citet{wang2022multi,sun2024structure,zhang2024structure} or contrastive learning \citep{wang2025s}.
Some studies have also leveraged pretrained protein language models to improve structure models~\citep{wu2023integration, zheng2024ccpl}.

\noindent \textbf{Knowledge-Enhanced Models}: Beyond large protein sequence datasets, information in other formats can further enhance a model’s understanding of proteins in the training stage. 
OntoProtein \citep{zhang2022ontoprotein} and KeAP \citep{zhou2023protein} incorporated knowledge graphs data during training by additional MLM objectives and/or contrastive learning to inject factual biological knowledge into the pre-trained \proteinllms.
ProteinBERT \citep{brandes2022proteinbert} performed dual-task learning during pretraining to learn both protein sequence modeling and Gene Ontology (GO) annotation prediction. It utilized a specialized BERT architecture with parallel input pathways for sequences and annotations.
To leverage the rich information in textual descriptions or other modalities, ProteinCLIP \citep{wu2024proteinclip} and MolBind \citep{xiao2024molbind} applied contrastive learning between protein sequences and textual descriptions and/or molecular to learn improved embeddings.

\subsection{Protein Description and Annotation Models}

The previously mentioned models have primarily focused on learning protein representations and utilizing them for classification, regression, or 3D structure folding tasks. To enhance expressiveness and understanding, more recent models have been trained on both protein sequences and textual data, allowing them to integrate NLP capabilities with protein representation learning \citep{wang2023instructprotein, liu2024prott3, zhuo2024protllm,jin2024prollm}.
\citet{xu2022protranslator} proposed ProTranslator, a bilingual translation framework between protein sequences and GO functions with textual descriptions. ProTranslator encoded and aligned the textual definitions of GO functions and protein sequences within the same low-dimensional space, facilitating the annotation of novel GO functions and the generation of textual descriptions for proteins. BioTranslator \citep{xu2023multilingual} further improved ProTranslator by extending the bilingual framework to a multilingual translation framework, embedding text and multiple biomedical modalities into a shared space.
ProtST~\cite{xu2023protst} was a framework designed to jointly learn from protein sequences and their associated biomedical text descriptions. It integrated protein language models (e.g., ESM or ProtBERT) with biomedical language models (e.g., PubMedBERT) to fuse sequence and text information through pre-training tasks. Prot2Text~\citep{abdine2024prot2text} combined ESM-2 with a structure encoder (RGCN) and extended function prediction from categorical classification to free-text descriptions. BioT5 and BioT5+~\citep{pei2023biot5,pei2024biot5+} further unified molecular information within a more comprehensive training framework.

There have also been several interactive LLMs for protein understanding. These models enhanced pretrained LLMs with protein comprehension by integrating a protein processing module~\citep{wu2024structure, wang2024protchatgpt,wang2024long}. For instance, ProteinChat~\citep{guo2023proteinchat} allowed users to input protein structures and query them using texts. ProteinGPT~\citep{xiao2024proteingpt} extended this capability by supporting both protein sequences and structures as inputs. In these models, protein data were processed through \proteinllms to generate embeddings, which were then projected to the natural language embedding space. The backbone LLMs integrated these adapted embeddings with user’s queries to produce meaningful answers.

\section{LLM Methods for Protein Engineering, Generation and Translation}
\label{sec:llm_generation}

Protein engineering and generation aims to design protein sequences with desired attributes (e.g. structures and properties). Given the desired attributes \(T\) and reference protein sequence \(\mathcal{S}\) (optional), the model is expected to output a protein sequence \(\mathcal{S}'\) with desired attributes. Key tasks include:

\noindent \textit{I. Protein Engineering:} 
\( f_\theta: (\mathcal{S}, T) \rightarrow \mathcal{S}' \) modifies protein \(\mathcal{S}\) toward the desired attributes \(T\), yielding the engineered protein \(\mathcal{S}'\).

\noindent \textit{II. Protein Generation:} 
\( f_\theta: (T, R) \rightarrow \mathcal{P} \) generates proteins with attributes \(T\) by sampling from the protein space using random seeds \(R\).

\noindent \textit{III. Protein Translation:} \( f_\theta: (\mathcal{P},T) \rightarrow \mathcal{P'} \) translates a protein \( \mathcal{P} \) into an alternative representation \( \mathcal{P}' \)  based on the target translation parameters \( T \).

\subsection{Protein Engineering Models}

ProteinDT~\cite{liu2023text} is a multimodal protein design framework that robustly integrates textual protein knowledge with sequence-based generative modeling. ProteinDT employs contrastive alignment and a facilitator module, enabling zero-shot text-to-protein generation and editing. 
Meanwhile, PLMeAE~\cite{plmeae} is a closed-loop protein engineering framework that integrates protein language models with an automated biofoundry within a Design-Build-Test-Learn cycle.
Furthermore, Toursynbio~\cite{shen2024toursynbio} introduces an agent that is capable of facilitating the modification and engineering of wet lab proteins.

\subsection{Protein Generation Models}

Protein generation models are designed to create novel protein sequences for specific engineering applications, often leveraging large-scale datasets of existing proteins with known amino acid sequences and properties. These models typically employ decoder-based architectures to generate functional protein sequences conditioned on various biological annotations. For example, ProGen~\cite{madani2023large} is a GPT-based generative protein engineering model that treats protein engineering as an unsupervised sequence generation process, and generates functional protein sequences conditioned on annotations like molecular function or taxonomy. The model is trained on diverse, non-redundant protein sequences from databases such as UniProt and Pfam, utilizing associated tags for conditional generation. ProtGPT2~\cite{ferruz2022protgpt2} is another model that generates de novo protein sequences with natural amino acid compositions using autoregressive modeling. In particular, they noticed that the generated sequences could explore a few uncharted areas of the protein sequence space. 
ProGen2~\cite{nijkamp2023progen2} is an extended version of ProGen, featuring a larger model size and a more extensive training dataset to enhance sequence diversity.  
Notably, ProGen2 can predict protein fitness without requiring additional fine-tuning. Recently, ProLLaMA~\citep{lv2024prollama} proposed a multi-task protein language model to handle both protein sequence generation and protein understanding tasks. Built on LLaMA2, ProLLaMA introduces a two-stage training framework: (1) continued pre-training on protein sequences, and (2) instruction tuning with a 13-million-sample dataset for multitasking capabilities.

Beyond conventional decoder-based approaches, Ankh~\citep{elnaggar2023ankhoptimizedproteinlanguage} employs an encoder-decoder architecture that optimizes efficiency by reducing parameters while maintaining high-quality protein generation. PAAG~\citep{yuan2024annotation} is another encoder-decoder architecture which focuses on the alignment between textual annotations and protein sequences at multiple levels before generating new sequences.  
Pinal~\citep{dai2024toward} does not directly generate protein sequences from text. Instead, it first constrains the protein design space by generating structure tokens, then predicts sequences based on those constraints to improve foldability and function alignment.

While many of these models are designed for general protein generation, some focus on specialized applications such as antibody design. IgLM~\cite{shuai2023iglm} employs autoregressive sequence generation conditioned on an antibody's sequence chain type and species of origin. 
As a further step, PALM-H3~\cite{he2024novo} specifically targets SARS-CoV-2 antibody generation, highlighting how protein generation language models can be tailored for highly specific protein design tasks.

\subsection{Protein Translation Models}

Protein translation models are specifically developed to handle tasks that require translating between different protein representations, which could be helpful in protein design. 

ProstT5~\cite{heinzinger2023bilingual} addresses the task of simultaneously modeling the dual nature of proteins — their linear one-dimensional (1D) sequences and three-dimensional (3D) structures — using a bilingual language model based on T5~\cite{raffel2020exploring} and ProtT5~\cite{pokharel2022improving}. It extracts features and patterns from both the sequence and the structure data 
Fold2Seq~\cite{cao2021fold2seq} is another model that learns structure-sequence relationships of proteins. 
The model could guide designs of protein sequences conditioned on desired structural folds. Recently, ProtAgents \cite{ghafarollahi2024protagents}, a multiagent framework, has been proposed to handle 1D sequence generation and 3D fold generation simultaneously. LM-DESIGN~\citep{10.5555/3618408.3620189} is a method for reprogramming protein language models (pLMs) to design protein sequences for given structural folds.

\section{Datasets}
\label{sec:dataset}
Datasets are crucial for training and evaluating \proteinllms. They are categorized into pretraining datasets, comprising unlabeled protein sequences for self-supervised learning, and benchmark datasets, which contain labeled sequences for supervised fine-tuning and evaluation on specific biological tasks.

\subsection{Pretraining Datasets}

\smallskip \noindent \textbf{UniProtKB}: A comprehensive protein sequence and annotation database composed of two main components: \textit{Swiss-Prot}~\cite{boutet2016uniprotkb}, a manually curated, high-quality dataset with reliable annotations and \textit{TrEMBL}~\cite{m1999edittotrembl}, an automatically annotated dataset providing broader coverage. 

\smallskip \noindent \textbf{UniRef Clusters}~\cite{suzek2015uniref}: A collection of clustered protein sequences designed to reduce data redundancy and improve computational efficiency. Provided by the UniProt database, UniRef is organized into three hierarchical levels: UniRef100, UniRef90, and UniRef50. UniRef100 contains a non-redundant set of all UniProt protein sequences where the latter two are created by clustering sequences with at least 90\% and 50\% sequence identity.

\smallskip \noindent \textbf{Pfam}~\cite{finn2006pfam}: A database of protein families and domains widely used for annotation and analysis of protein sequences. Each Pfam entry represents a group of related protein sequences defined by a multiple sequence alignment and a corresponding profile hidden Markov model (HMM). It provides insights into protein structure, function, and evolution, helping researchers identify conserved domains, predict functions, and classify proteins across organisms.

\smallskip \noindent \textbf{PDB}~\cite{bank1971crystallography}: The Protein Data Bank is a repository for the 3D structural data of large biological molecules, such as proteins and nucleic acids. It provides valuable resources for understanding the structural aspects of proteins, which can be beneficial for training models that incorporate structural information.

\smallskip \noindent \textbf{AlphaFoldDB}~\cite{tunyasuvunakool2021highly}: The AlphaFold Protein Structure Database offers predicted protein structures generated by the AlphaFold model containing over 200 million entries.

\subsection{Benchmark Datasets}

\smallskip \noindent \textbf{CASP}~\cite{kryshtafovych2019critical}: Critical Assessment of Structure Prediction is a biennial competition that evaluates methods for protein structure prediction. Participants predict 3D structures of proteins from their sequences, compared against experimental results.

\smallskip \noindent \textbf{ProteinGym}~\cite{notin2024proteingym}: A large-scale benchmark platform for protein design and fitness prediction. It includes over 250 Deep Mutational Scanning (DMS) assays, encompassing millions of mutated protein sequences, and curated clinical datasets with expert annotations. By integrating zero-shot and supervised evaluation frameworks, ProteinGym allows systematic comparison of over 70 machine learning models. It provides standardized metrics for tasks like mutation effect prediction and protein design, fostering innovation in computational biology and protein engineering.

\smallskip \noindent \textbf{TAPE}~\cite{rao2019evaluating}: A benchmark designed to evaluate protein sequence embeddings in biologically relevant tasks using machine learning. It includes five tasks covering structure prediction, evolutionary understanding, and protein engineering. TAPE leverages self-supervised learning, enabling models to learn from unlabeled protein sequences, and offers standardized datasets and metrics for systematic comparisons. It aims to advance protein representation learning by addressing gaps in generalization and real-world applicability.

\smallskip \noindent \textbf{PEER}~\cite{xu2022peer}: A comprehensive and multi-task benchmark designed to evaluate protein sequence understanding. It includes tasks such as protein function prediction, localization prediction, structure prediction, protein-protein interaction prediction, and protein-ligand interaction prediction. 

\smallskip \noindent \textbf{ProteinLMBench}~\cite{shen2024fine}: A benchmark dataset comprising 944 manually verified multiple-choice questions aimed at assessing the protein understanding capabilities of LLMs. It incorporates protein-related details and sequences in multiple languages, setting a new standard for evaluating LLMs’ abilities in protein comprehension.

\section{Evaluation Metrics}
\label{sec:eval}
Comprehensive evaluation is essential for applying \proteinllms, which are assessed on tasks like structure prediction, function prediction, and sequence generation. Appendix~\ref{sec:eval} provides detailed descriptions of structure and function prediction metrics, as well as sequence generation metrics for generative \proteinllms.

\subsection{Structure Prediction Metrics}
\texttt{Root Mean Square Deviation (RMSD)} measures the distance between predicted and actual atomic coordinates, with lower values indicating better accuracy~\cite{li2013difficulty}. \texttt{Global Distance Test (GDT-TS)} calculates the percentage of alpha-carbon atoms within 1, 2, 4, and 8 \r{A} thresholds, reflecting structural similarity~\cite{zemla2003lga}. \texttt{Template Modeling (TM) Score} evaluates global structural similarity (scores between 0 and 1) via 
\begin{gather}
\text{TM}=\max \left[ \frac{1}{L_{\text{tgt}}} \sum_{i}^{L_{\text{com}}} \frac{1}{1+\left( \frac{d_{i}}{\scriptscriptstyle d_{0}(L_{\text{tgt}})} \right)^{2}} \right], \\
d_0(L_{\text{tgt}}) = 1.24 \sqrt[3]{L_{\text{tgt}} - 15} - 1.8.
\end{gather}
\texttt{Local Distance Difference Test (lDDT)} quantifies local accuracy by comparing interatomic distances~\cite{mariani2013lddt}, and \texttt{Predicted Local Distance Difference Test (pLDDT)} provides per-residue confidence scores (0–100) without a reference structure, as used in AlphaFold~\cite{guo2022alphafold2,jumper2021highly}.

\subsection{Function Prediction Metrics}

Protein function prediction determines biological roles, including biomolecular interactions~\cite{radivojac2013large}. Machine learning metrics include classification measures (precision, recall, F-1 score, accuracy, AUC) and generative metrics such as BLEU~\cite{papineni2002bleu}, ROUGE-L~\cite{lin2004rouge}, and METEOR~\cite{banerjee2005meteor}. These evaluation methods offer quantitative benchmarks crucial for model validation and biological inference.

\texttt{Subcellular Localization} predicts proteins' cellular positions to infer functions~\cite{briesemeister2010yloc,holm2020dali}. \texttt{Homology Detection} identifies evolutionary relationships using sequence alignment methods like BLAST~\cite{altschul1990basic} or deep learning approaches such as TM-vec~\cite{hamamsy2024protein}. \texttt{Stability} and \texttt{Solubility} assessments evaluate whether a protein can function effectively in its environment~\cite{cheng2006prediction,hebditch2017protein}, while \texttt{Mutation Effect Prediction} gauges the impact of amino acid changes on protein properties~\cite{mansoor2022accurate}. These integrative metrics underpin the development of robust protein prediction systems and support advancements in drug design and molecular biology.

\section{Conclusion and Future Work}
\label{sec:conclusion}

This survey provides a comprehensive overview of Protein Large Language Models, highlighting their architectures, datasets, evaluation, and applications. These works represent significant advancements in protein science and offer innovative approaches to protein analysis and design. In addition to these advancements, several challenges remain to be solved in the future.

\smallskip \noindent \textbf{Protein Dynamics.} AlphaFold~\cite{jumper2021highly} has been shown to provide accurate static 3D structures. However, proteins are naturally dynamic molecules with various conformations~\cite{ohnuki2024integration}. Although several works incorporate 3D structures into LLMs, the conformational dynamics of proteins have not yet been considered. Since conformational dynamics are highly related to the transporter functions of proteins, it would benefit the model to include protein dynamics.

\smallskip \noindent \textbf{Combination with Single-cell Data.} Recently, single-cell proteomics sequencing technology~\cite{li2024scprotein,liu2024geneverse,bennett2023single} has attracted extensive attention in the field of biology, which can help us understand the pathways in specific cells. Since LLMs have shown effectiveness in understanding both proteins and single-cell data, they can be extended to learn from single-cell proteomics data in the future.

\smallskip \noindent \textbf{Towards Biological Applications.} Although several biological applications have been studied in recent works, a range of detailed and complex problems remain unsolved, including protein-ligand interaction learning~\cite{koh2024physicochemical}, cryptic pocket identification~\cite{ge2024exploring}, and rational ligand generation~\cite{li2024deep}. These applications require extensive and diverse domain knowledge of proteins and their related fields. We believe LLMs have the potential to incorporate and utilize more domain knowledge to solve these problems.

\smallskip \noindent \textbf{Interpretability.} In addition to effectiveness, interpretability is also of strong significance for trustworthy models~\cite{huang2024trustllm}. Previous language models for proteins~\cite{gu2023hierarchical,vecchietti2024recent} have provided extensive case studies, such as key residue analysis, which could be challenging for large-scale and closed-source models. To improve interpretability, InterPLM~\cite{simon2024interplm} employs sparse autoencoders to extract biologically meaningful features from \proteinllms, revealing their alignment with known biological concepts. Inspired by this, we should design prompts to enhance the interpretability of \proteinllms for reliable outputs.

\newpage
\section*{Limitations}

This survey primarily focuses on \proteinllms. We acknowledge that the study of protein interactions with other molecules (e.g., DNA, RNA) in the inter-molecular domain is a broad and valuable field worth reviewing. Given its vast scope, we do not extensively cover it in this survey, and instead focus on \proteinllms centered on proteins themselves. In the future, we may either expand our review to include these areas or write a separate survey specifically dedicated to this domain, providing more comprehensive coverage for researchers.

\bibliography{main}
\newpage
\appendix

\label{sec:appendix}

\begin{figure*}[htbp]
    \centering
    \includegraphics[width=0.95\textwidth]{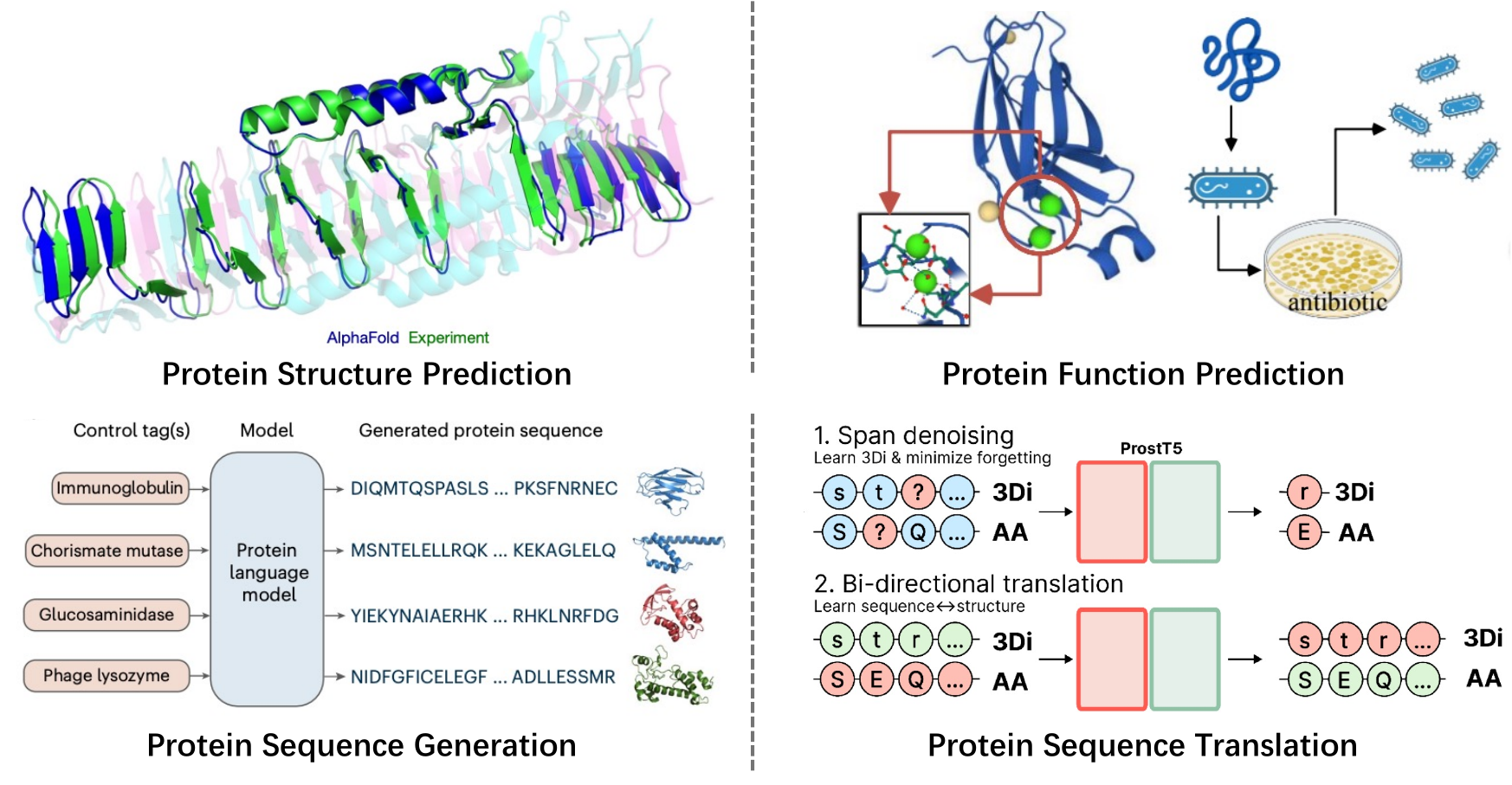}
    \caption{Illustrations on General Tasks of Protein Language Models.}
    \label{fig:task}
\end{figure*}

\section{Experimental Methods in Proteomics and Their Limitations}
\label{sec:experiment}

Traditional experimental techniques such as X-ray crystallography, nuclear magnetic resonance (NMR) spectroscopy, and cryo-electron microscopy (cryo-EM) in protein science have laid the foundation for studying protein structure and functions. However, computational approaches and also embrace the progress of AI development. This section briefly covers methods, which are essential for determining protein structures and functions.

\smallskip \noindent \textbf{X-ray Crystallography} is a widely utilized method for determining the 3D structures of proteins~\citep{jones1986using}. 
In this method, X-rays are directed at a crystallized sample, and the resulting diffraction patterns are analyzed to reveal the arrangement of atoms within the crystal. 
This process provides detailed insights into the protein's electron density and overall structure. 
However, crystallization can be challenging, especially for large, flexible, or membrane-associated proteins. The technique typically offers a static snapshot of the protein, which may not fully capture its dynamic nature in solution.
Advancements in AI have led to the development of structure prediction tools like AlphaFold~\citep{jumper2021highly} and RoseTTAFold~\citep{baek2021accurate}. For instance, the crystal structure of the KlNmd4 protein is predicted to consist of a single PIN domain~\citep{barbarin2021artificial}. The study demonstrates that the high-quality models significantly accelerate the determination of KlNmd4's structure, while existing models fail to achieve similar results.

\smallskip \noindent \textbf{Nuclear Magnetic Resonance (NMR) Spectroscopy} is a non-destructive technique for determining the structure, dynamics, and interactions of molecules at the atomic level under near-physiological conditions~\citep{shukla2023biomolecular}. 
It provides 3D structural data of proteins in solution and captures real-time dynamics, making it highly effective for studying protein flexibility and weak protein-ligand interactions. NMR exploits the magnetic properties of atomic nuclei (e.g., hydrogen nuclei in proteins) to provide detailed information about the local chemical environment.

With the development of AI, deep learning methods are more and more promising to advance the reconstruction of sparsely sampled data in NMR spectroscopy, particularly in the context of non-uniform sampling. The input data typically consists of sparsely sampled NMR spectra, while the output is the fully sampled spectrum, reconstructed either in the time~\citep{hansen2019using,karunanithy2021fid} or frequency domain~\citep{qu2020accelerated,luo2020fast}. For time-domain reconstructions, neural networks effectively predict the missing data points. In frequency-domain reconstructions, they excel at removing artifacts caused by sparse and non-Nyquist sampling. Studies across various research groups have consistently demonstrated the high accuracy of DNN-based reconstructions, even under conditions of extremely sparse sampling, highlighting the potential of deep learning to enhance data acquisition and analysis in NMR.

However, NMR has limited size range: NMR is mostly suitable for proteins smaller than ~30–50 kDa (larger proteins become challenging due to signal overlap). Protein sample preparation and data collection can also be expensive and take weeks to months.

\smallskip \noindent \textbf{Cryo-EM} is a structural biology technique that enables the direct observation of conformational heterogeneity in individual dynamic macromolecules~\citep{lyumkis2019challenges}. Researchers aim to reconstruct high-resolution 3D structural landscapes from numerous 2D observed projections, which may represent different conformational states. However, the cryo-EM reconstruction task is challenging because each particle's pose is unknown during imaging.
Recently, deep learning methods have demonstrated powerful capabilities in representing heterogeneity within datasets by mapping them onto nonlinear manifold embeddings. On the one hand, CryoDRGN~\citep{zhong2021cryodrgn} is a pioneering work that captures this heterogeneity by employing variational autoencoders (VAEs) to map the data into a low-dimensional latent space. A generative decoder then reconstructs a 3D volume from a sampled point in this latent space.
CryoGAN~\citep{gupta2021cryogan} introduces an entirely new possibility to learn to reconstruct in a distributional sense with a generative adversarial framework. Because of its likelihood-free nature, CryoGAN does not require any additional processing steps such as pose estimation and can be directly applied to cryo-EM measurements. This greatly simplifies the reconstruction procedure. On the other hand, E2gmm~\citep{chen2021deep} models the 3D structure using a set of Gaussians to automatically resolve the structural heterogeneity, whereas 3DFlex~\citep{punjani20233dflex} employs a neural network to fit the 3D displacement field of each particle by concurrently exploring its deformation field and refining a canonical density. More recently, CryoSTAR~\citep{li2024cryostar} resolves continuous conformational heterogeneity by constructing reasonable coarse-grained models, meanwhile, density maps are also estimated for different conformations. It meticulously preserves local structures, minimizes erroneous solutions, and ultimately achieves enhanced, accelerated convergence. Overall, the current trend is to incorporate atomic information to better activate deep models, aiming for more precise 3D structures that better comply with natural laws.

\clearpage

\newcommand{\centeredrottable}[1]{%
    \begin{center}
    \vspace*{\textheight}  
    \begin{table*}[!h]
    \centering
    #1
    \end{table*}
    \end{center}
}

\begin{table*}[htbp]
  \centering
  \caption{LLM Methods for Protein Understanding and Prediction: Protein Sequence Models}
  \label{tab:protein-engineering-generation-1}
  \small
  \renewcommand{\tabularxcolumn}[1]{m{#1}}
  \begin{tabularx}{\linewidth}{%
      >{\centering\arraybackslash}m{2.5cm}  
      >{\centering\arraybackslash}m{0.7cm}  
      >{\centering\arraybackslash}m{1.5cm}  
      >{\centering\arraybackslash}m{5cm}  
      >{\centering\arraybackslash}X}         
    \toprule
    \textbf{Model} & \textbf{Time} & \textbf{Base Model} & \textbf{Dataset} & \textbf{Keywords} \\
    \midrule
    UniRep~\citep{alley2019unified} & 2019 & BiLSTM & UniRef50 & Representation learning, Stability prediction, Functional effects of mutations \\
    \midrule
    \citet{bepler2019learning} & 2019 & BiLSTM & SCOPe ASTRAL, Pfam, PDB, TOPCONS, CASP12 & Structural property prediction, Soft symmetric alignment, Transmembrane \\ 
    \midrule
    MuPIPR \citep{zhou2020mutation} & 2020 & BiLSTM & STRING, PDB, SKP1402m, SKP1102s & Protein–Protein Interactions (PPI), binding affinity, buried surface area \\ 
    \midrule
    CSCS \citep{hie2021learning} & 2020 & BiLSTM & IRD,LANL HIV database,ViPR,NCBI Virus,GISAID  & Viral escape patterns, Constrained Semantic Change Search \\ 
    \midrule
    ProtTrans~\citep{elnaggar2021prottrans} & 2021 & Transformer-XL, XLNet, BERT, Albert, Electra, T5 & UniRef, BFD & Protein secondary structure, sub-cellular localization, membrane vs. water-soluble \\ 
    \midrule
    ESM-1b~\citep{rives2021biological} & 2021 & Transformer & Uniparc & Large-scale pretraining, protein structure, functional effects of mutations \\ 
    \midrule
    ESM-1v~\citep{meier2021language} & 2021 & ESM-1b & Uniref90 & Functional effects of mutations, zero-shot prediction \\ 
    \midrule
    ESM-2, ESMFold~\citep{lin2023evolutionary} & 2023 & Transformer & UniRef, PDB, CAMEO, CASP14, MGnify, trRosetta Dataset & Atom-level resolution structure prediction \\ 
    \midrule
    AminoBERT~\citep{chowdhury2022single} & 2022 & BERT & ProteinNet12, SCOPe ASTRAL & Single-sequence protein structure prediction \\ 
    \midrule
    TCR-BERT~\citep{wu2024tcr} & 2021 & BERT & VDJdb, PIRD, LCMV dataset & TCR–antigen binding \\ 
    \midrule
    MSA Transformer~\citep{rao2021msa} & 2021 & Transformer & UniRef50, UniClust30, CASP13, CAMEO & Multiple sequence alignment, evolutionary relationships \\ 
    \midrule
    Tranception~\citep{notin2022tranception} & 2022 & Transformer & UniRef & Homologous sequences retrieval, fitness prediction \\ 
    \midrule
    XTrimoPGLM~\citep{chen2024xtrimopglm} & 2024 & Transformer & UniRef90, ColabFoldDB, UniProt, AlphaFold Database, PDB & 100B parameters, Unified Protein Language Model \\ 
    \midrule
    TurNuP \citep{kroll2023turnover} & 2022 & ESM-1b &BRENDA, UniProt, Sabio-RK &Turnover number predictions, Differential Reaction Fingerprints \\ 
    \midrule
    CLEAN \citep{yu2023enzyme} & 2023 & ESM-1b &UniProt,SwissProt   & Contrastive Learning, Enzymatic function prediction \\ 
    \midrule
    DeepTMP \citep{lin2023deep} & 2023 & ESM-MSA-1b & PDB, PDBTM, UniRef30, BFD & Transfer learning, Transmembrane protein complexes,Inter-chain Contact Prediction \\ 
    \midrule
    vcMSA~\citep{mcwhite2023leveraging} & 2023 & ProtT5-XL-UniRef50 & Quantest2, HOMSTRAD, UniRef50 & MSA identification, Reciprocal Best Hits\\ 
    \midrule
    Poet~\citep{truong2023poet} & 2023 & Transformer  & UniRef50, UniRef100, ProteinGym & Homologous Sequences, Retrieval-augmented LM \\ 
    \bottomrule
  \end{tabularx}
\end{table*}

\clearpage

\begin{table*}[htbp]
  \centering
  \caption{LLM Methods for Protein Understanding and Prediction: Structure-Integrated and Knowledge-Enhanced Models}
  \label{tab:PUP_structure_knowledge}
  \small
  \renewcommand{\tabularxcolumn}[1]{m{#1}}
  \begin{tabularx}{\linewidth}{%
      >{\centering\arraybackslash}m{2.5cm}  
      >{\centering\arraybackslash}m{0.7cm}  
      >{\centering\arraybackslash}m{2.0cm}  
      >{\centering\arraybackslash}m{4cm}  
      >{\centering\arraybackslash}X}         
    \toprule
    \textbf{Model} & \textbf{Time} & \textbf{Base Model} & \textbf{Dataset} & \textbf{Keywords} \\
    \midrule
    ProteinBERT~\citep{brandes2022proteinbert} & 2022 & BERT & UniRef90, TAPE & GO annotations, protein structures, post-translational modifications, biophysical properties \\ 
    \midrule
    OntoProtein~\citep{zhang2022ontoprotein} & 2022 & ProtBert, Bert & ProteinKG25, UniRef100, TAPE, STRING, SHS27k, SHS148k & Knowledge graphs, gene ontology, PPI, structure prediction \\ 
    \midrule
    ProteinCLIP~\citep{wu2024proteinclip} & 2024 & ESM2, ProtT5, Text-Embedding-3-Large & UniProt & Contrastive learning, PPI, homology identification \\ 
    \midrule
    SaProt~\citep{su2023saprot} & 2023 & ESM2 & AlphaFoldDB, UniProt, ProteinGym, ClinVar, thermostability, metal ion binding, DeepLoc, TAPE, PEER, FLIP, PDB & Structure-aware vocabulary, Foldseek \\ 
    \midrule
    ESM-GearNet ~\citep{zhang2023systematic} & 2023 & GVP, GearNet, CDConv &  AlphaFold Database, GO~\citep{gligorijevic2021structure}, Atom3D & Structural encoders for protein modeling \\ 
    \midrule
    SES-Adapter\citep{tan2024simple} & 2024 & ESM2, ProtBert, ProtT5, Ankh & GO \citep{gligorijevic2021structure} & Parameter-Efficient Fine-Tuning, Structure Representation \\ 
        \midrule
     PromptProtein ~\citep{wang2022multi}, &2023 & Transformer & UniRef50, PDB, STRING, GO~\citep{gligorijevic2021structure}& Prompt Learning, Multi-level of structures\\ 
          \midrule
     SI-pLMs ~\citep{sun2024structure} & 2024 & BERT & Pfam, PDB, AlphaFold Database & Variant Effect Prediction, Structural Information\\ 
     \midrule
     \citet{zhang2024structure} & 2024 & ESM-2 & SCOPe, GO and EC \citep{gligorijevic2021structure}, Swiss-Prot & Remote Homology Detection, Structural Information \\ 
         \midrule
   S-plm~\citep{wang2025s} & 2025 &ESM3, discrete diffusion & BPTI, RMSD, Apo/holo, Fold-switch, ATLAS  & Contrastive Learning, Structural Information \\ 
    \midrule
    \citet{wu2023integration} & 2023 & ESM-2, MSA-Transformer, GVP-GNN, EGNN, SE(3)-Transformer, Schnet, DimeNet& CASP, DB5.5, DIPS, PDBbind & Geometric Deep Learning \\ 
    \midrule 
    CCPL~\citep{zheng2024ccpl} & 2023-2024 & GVP-GNN, ESM-2 & PDB,  AlphaFoldDB, ProteinGym, trRosetta, CASP14, CATH, Ts 50\&Ts500 & Contrastive Learning, Structure-
Sequence Pairing \\ 
    \midrule
    KeAP~\citep{zhou2023protein} & 2023 & ProteinKG25 & ProteinNet,TAPE  & Knowledge Graph, Contrastive Learning \\ 
    \midrule
    MolBind~\citep{xiao2024molbind} & 2024 & SciBERT, GIN, Uni-Mol  & MolBind-M4, CASF-2016  & Contrastive Learning, Protein-text-molecule Alignment \\ 
    \bottomrule
  \end{tabularx}
\end{table*}

\clearpage

\begin{table*}[htbp]
  \centering
  \caption{LLM Methods for Protein Understanding and Prediction: Protein Description and Annotation Models}
  \label{tab:PUP_description_annotation}
  \small
  \renewcommand{\tabularxcolumn}[1]{m{#1}}
  \begin{tabularx}{\linewidth}{%
      >{\centering\arraybackslash}m{2.5cm}  
      >{\centering\arraybackslash}m{0.7cm}  
      >{\centering\arraybackslash}m{2.0cm}  
      >{\centering\arraybackslash}m{4cm}  
      >{\centering\arraybackslash}X}         
    \toprule
    \textbf{Model} & \textbf{Time} & \textbf{Base Model} & \textbf{Dataset} & \textbf{Keywords} \\
    \midrule
    ProtST~\cite{xu2023protst} & 2023 & ProtBert, PubMedBERT, etc.& ProtDescribe & Multimodal learning, protein function annotation, zero-shot text-to-protein retrieval \\ 
    \midrule
    ProtChatGPT~\cite{wang2024protchatgpt} & 2024 & ESM-1b, Transformer & PDB-QA, ProteinKG25 &Protein Q\&A, cross-modal protein retrieval, qualitative dialogs \\ 
    \midrule
    ProteinChat~\cite{guo2023proteinchat} & 2023 & ESM-IF1, Vicuna-13B & RCSB-PDB Protein Description & Interactive protein inquiries, automated protein understanding \\ 
       \midrule
    Prot2Text~\citep{abdine2024prot2text} & 2024 & RGCN, ESM2, GPT2  & SwissProt & Multimodality, textual function prediction \\ 
    \midrule
    ProTranslator~\citep{xu2022protranslator} & 2022 & DeepGOCNN, Transformer & CAFA3, SwissProt, GOA, Reactome, KEGG, MSigDB & Function annotation based on text description, text description generation \\ 
     \midrule
    BioTranslator~\citep{xu2023multilingual} & 2023 & PubMedBERT & GOA, Swiss-Prot, CAFA3, STRING, GeneCards, Tabula Muris, Tabula Sapiens, Tabula Microcebus, GDSC, STITCH, Monarch Initiative, Reactome & Multimodality, text-to-bio-identity translation \\ 
    \midrule
    BioT5~\citep{pei2023biot5} & 2023 & T5 & ZINC20, UniRef, C4, PubMed articles, PubChem, ChEBI20, SwissProt, MoleculeNet, PEER, BindingDB, BioSNAP, HPRD, Yeast PPI dataset & SELFIES-based molecular representation, wrapped text for bio-entities \\ 
    \midrule
    BioT5+~\citep{pei2024biot5+} & 2024 & T5 & MoleculeNet, ChEBI-20, PEER, BioSNAP, BindingDB & Multi-task instruction tuning, Molecular\\
    \midrule
    ProLLaMA~\citep{lv2024prollama} & 2024 & LLaMA2 & UniRef, InterPro & Instruction understanding, protein understanding and generation \\ 
     \midrule
    ProteinGPT~\cite{xiao2024proteingpt} & 2024 & ESM-2, ESM-IF1, Vicuna, LLaMA-2, LLaMA-3 & ProteinQA & Multimodal, interactive protein Q\&A \\ 
    \midrule
    ProLLM\citep{jin2024prollm} & 2024 & Flan-T5-large & Human, STRING, Mol-Instructions & Chain-of-Thought, PPI \\
    \bottomrule
  \end{tabularx}
\end{table*}

\clearpage
\begin{table*}[htbp]
  \centering
  \caption{LLM Methods for Protein Engineering, Generation and Translation}
  \label{tab:generative-llm-protein}
  \small
  \renewcommand{\tabularxcolumn}[1]{m{#1}}
  \begin{tabularx}{\linewidth}{%
      >{\centering\arraybackslash}m{2.5cm}  
      >{\centering\arraybackslash}m{1cm}  
      >{\centering\arraybackslash}m{2.5cm}  
      >{\centering\arraybackslash}m{3cm}  
      >{\centering\arraybackslash}X}         
    \toprule
    \textbf{Model} & \textbf{Time} & \textbf{Base Model} & \textbf{Dataset} & \textbf{Keywords} \\
    \midrule
    ProGen~\cite{madani2023large} & 2020 & Transformer & UniParc, UniProtKB, Swiss-Prot, TrEMBL & Controllable protein generation, de novo protein design \\
     \midrule
    ProGen2~\cite{nijkamp2023progen2} & 2022 & Autoregressive & UniRef50 & Protein generation, de novo protein design \\
     \midrule
    ProtGPT2~\cite{ferruz2022protgpt2} & 2022 & Autoregressive & UniRef50 & Autoregressive transformer, BPE tokenization, zero-shot protein generation \\
     \midrule
    ProLLaMA~\cite{lv2024prollama} & 2024 & LLaMA2 & UniRef50, InterPro & Multi-task, instruction tuning \\
     \midrule
    IgLM~\cite{shuai2023iglm} & 2023 & GPT-style Transformer & OAS Training Data, Thera-SAbDab & Infilling, conditioned generation, controllable diversity \\
     \midrule
    PALM-H3~\cite{he2024novo} & 2024 & ESM2, RoFormer & Observed Antibody Space, CoV-AbDab, BioMap & Strong generalization to novel proteins, interpretability, antibody \\
     \midrule
    ProstT5~\cite{heinzinger2023bilingual} & 2023 & T5, ProtT5 & 3Di from AlphaFoldDB, CASP12/14, NetSurfP2.0 & Bilingual LM, Foldseek, inverse folding \\
     \midrule
    Fold2Seq~\cite{cao2021fold2seq} & 2021 & Transformer & CATH 4.2 & Inverse protein design, fold-level representation \\
     \midrule
    Ankh~\cite{elnaggar2023ankhoptimizedproteinlanguage} & 2023 & T5 & UniRef50, CASP12/14, NetSurfP-2.0, DeepSF, etc & Contact prediction, secondary structure, fold classification, efficiency \\
     \midrule
    ProteinDT~\cite{liu2023text} & 2023 & ProtBert, SciBERT, ProteinDiff, T5 & SwissProtCLAP & Multimodal learning, text-to-protein generation, autoregressive \\
    \midrule
    PLMeAE~\cite{plmeae} & 2025 & ESM-2 & GB1, UBC9 dataset, Ubiquitin  & Protein engineering, automatic biofoundry\\
    \midrule
    ESM-IF~\citep{hsu2022learning} & 2022 & GVP, GNN, Transformer & UniRef50, CATH  & Inverse folding, AlphaFold2 augmented dataset \\
    \midrule
    ESM-3~\citep{hayes2025simulating} & 2024 & Transformer & UniProt, PDB, AlphaFoldDB, Pfam, InterPro, MGnify, JGI, GO Consortium & Multimodal Learning, Evolutionary Simulation\\
    \midrule
    PAAG~\citep{yuan2024annotation} & 2024 & ProtBERT, SciBERT & ProtAnnotation & Text alignment, annotation\\
        \midrule
    Pinal~\citep{dai2024toward} & 2024 & T2struct, SaProt-T & SwissProt, UniRef50-ProTrek & Multi-step, functional labels\\
    \midrule
    ProtAgents~\citep{ghafarollahi2024protagents} & 2024 & GPT-4, Chroma,  OmegaFold & 
GPTProteinPretrained & Multi-agent, de novo protein design, protein folding\\
\midrule
Toursynbio~\citep{shen2024toursynbio} & 2024 & InternLM2-7B & ProteinLMDataset & Multi-modal, agent, interactive\\
\midrule
LM-DESIGN~\citep{10.5555/3618408.3620189} & 2024 & ESM-1b, ESM-2, ProteinMPNN & CATH 4.2, CATH 4.3, TS50, TS500 & De novo protein design, protein folding\\
    \bottomrule
  \end{tabularx}
\end{table*}

\clearpage

\begin{table*}[htbp]
  \centering
  \caption{Summary of Datasets for Protein Language Model}
  \label{tab:protein-datasets}
  \small
  \renewcommand{\tabularxcolumn}[1]{m{#1}}
  \begin{tabularx}{\linewidth}{%
      >{\centering\arraybackslash}m{2cm}  
      >{\centering\arraybackslash}m{4cm}    
      >{\centering\arraybackslash}m{1cm}  
      >{\centering\arraybackslash}m{1cm}    
      >{\centering\arraybackslash}X}         
    \toprule
    & \textbf{Dataset} & \textbf{Last Update} & \textbf{Scale} & \textbf{Keywords} \\
    \midrule
    \multirow{17}{*}{Pretraining} 
      & UniProtKB/Swiss-Prot~\cite{boutet2016uniprotkb} & 2025 & 573K & Manually curated, high-quality annotations, reviewed \\
      & UniProtKB/TrEMBL~\cite{m1999edittotrembl} & 2025 & 253M & Computationally annotated, unreviewed, automated predictions \\
      & UniRef Clusters~\cite{suzek2015uniref} & 2025 & $>$250M & Clustered sequences, reduced redundancy, hierarchical organization \\
      & Pfam~\cite{finn2006pfam} & 2024 & 22k & Protein families, HMMs, functional domains \\
      & PDB~\cite{bank1971crystallography} & 2025 & 231K & Protein structures, crystallography, molecular modeling \\
      & BFD~\cite{steinegger2018clustering} & 2021 & 2.5B & Massive protein database, sequence clustering, structure prediction \\
      & UniParc~\cite{bairoch2005universal} & 2025 & $>$250M & Non-redundant, protein sequence archive, database cross-referencing \\
      & PIR~\cite{barker2001protein} & 2025 & 513M & Protein sequence database, functional annotation, evolutionary classification \\
      & AlphaFoldDB~\cite{tunyasuvunakool2021highly} & 2025 & $>$200M & Predicted protein structures, deep learning, proteome coverage \\
     \midrule
    \multirow{13}{*}{Benchmark} 
      & CASP~\cite{kryshtafovych2019critical} & 2024 & N/A & Protein structure prediction, modeling competitions \\
      & ProteinGym~\cite{notin2024proteingym} & 2024 & 2.7M & Protein mutations, deep mutational scanning \\
      & TAPE~\cite{rao2019evaluating} & 2021 & $\sim$120K & Protein embeddings, sequence modeling \\
      & CATH~\cite{orengo1997cath} & 2024 & $>$150M & Structure classification, evolutionary relationships, domain hierarchy \\
      & PEER~\cite{xu2022peer} & 2022 & $>$60K & Protein understanding, multi-task benchmark, sequence evaluation \\
      & ExplorEnz~\cite{mcdonald2009explorenz} & 2025 & 8K & Enzyme classification, EC numbering, catalytic reactions \\
      & HIPPIE~\cite{schaefer2012hippie} & 2022 & 39K & Human protein interactions, network analysis \\
      & ProteinLMBench~\cite{shen2024fine} & 2024 & 893K & Protein language understanding, multiple-choice QA, model evaluation \\
    \bottomrule
  \end{tabularx}
\end{table*}

\clearpage

\section{Evaluation Metrics}
\label{sec:app_eval}
Comprehensive and accurate evaluation is essential for understanding and applying \proteinllms. Currently, these models are commonly assessed on tasks such as structure prediction, function prediction, and sequence generation.

\subsection{Structure Prediction Metrics}
Structure prediction evaluates how accurately a model predicts a protein’s three-dimensional structure from its sequence~\cite{kuhlman2019advances}. Common metrics include:

\smallskip\noindent \textit{\textbf{Root Mean Square Deviation (RMSD)}} measures the distance between the predicted and actual atomic coordinates. Lower RMSD indicates higher structural accuracy~\cite{li2013difficulty}.

\smallskip\noindent \textit{\textbf{Global Distance Test (GDT-TS)}} calculates the percentage of alpha-carbon atoms within thresholds (1, 2, 4, and 8 \r{A}) of the reference structure after iterative superimposition~\cite{zemla2003lga}.

GDT-TS usually uses thresholds of 1, 2, 4, and 8 \r{A}. The higher the GDT-TS score, the closer the predicted structure is to the reference structure.

\smallskip\noindent \textit{\textbf{Template Modeling (TM) Score}} evaluates the global structural similarity of proteins with values ranging from 0 to 1~\cite{zhang2004scoring}. 
\begin{gather}
\text{TM}=\max \left[ \frac{1}{L_{\text{tgt}}} \sum_{i}^{L_{\text{com}}} \frac{1}{1+\left( \frac{d_{i}}{\scriptscriptstyle d_{0}(L_{\text{tgt}})} \right)^{2}} \right], \\
d_0(L_{\text{tgt}}) = 1.24 \sqrt[3]{L_{\text{tgt}} - 15} - 1.8.
\end{gather}
Here, $L_{\text{tgt}}$ is the length of the target protein amino acid sequence. $L_{\text{com}}$ is the number of residues in the template and target structures. $d_i$ represents the distance between the $i$-th residue pair in the template structure and the target structure. Higher scores indicate closer similarity. 

\smallskip\noindent \textit{\textbf{lDDT}}, Local Distance Difference Test, evaluates the local accuracy of protein structure prediction by comparing distances between atom pairs in the predicted structures and those in the reference structures~\cite{mariani2013lddt}.

A distance is considered preserved if it falls within a specified threshold. lDDT is calculated as the proportion of preserved distances, with higher values indicating better local accuracy. 

\smallskip\noindent \textit{\textbf{pLDDT}}, Predicted Local Distance Difference Test, is a per-residue measure of local confidence~\cite{guo2022alphafold2}. pLDDT evaluates the local quality of the predicted structure without a reference structure. Its computation usually relies on models such as AlphaFold \cite{jumper2021highly}, which learns patterns from large-scale protein data. Scores range from 0 to 100, with higher scores indicating greater confidence and more accurate predictions.

\subsection{Function Prediction Metrics}
Protein function prediction aims to determine biological roles, including interactions with other biomolecules \cite{radivojac2013large}. The evaluation methods involve machine learning performance metrics and biomedical relevance validation.

Machine learning evaluation metrics can be categorized into classification task metrics and generative task metrics. For classification tasks, such as protein classification and interaction prediction, standard metrics can be adopted, such as precision, recall, F-1 scores, accuracy, and area under the curve (AUC). For generative tasks, such as question answering, evaluation is performed by measuring the alignment between the LLM's output and the ground truth using metrics such as BLEU \cite{papineni2002bleu}, ROUGE-L \cite{lin2004rouge}, and METEOR \cite{banerjee2005meteor}.

In addition to machine learning metrics, there are also biometric-related evaluation metrics:

\smallskip\noindent \textit{\textbf{Subcellular Localization}} refers to the specific location of proteins within a cell \cite{briesemeister2010yloc}. The location of a protein is closely related to the function it performs, so by predicting the subcellular localization of a protein, it is possible to speculate on the biological function it may have \cite{holm2020dali}.

\smallskip\noindent \textit{\textbf{Homology Detection}} aims to identify proteins that share an evolutionary relationship (homologous) with the target protein, usually reflected in similarities in  sequences, structure, and functions. Traditional methods such as BLAST \cite{altschul1990basic} 
perform sequence alignment to identify homologs by comparing the query sequence against a database.

Recent deep learning approaches such as TM-vec \cite{hamamsy2024protein} focus on structural similarity and generate vector representations of proteins.

\smallskip\noindent \textit{\textbf{Stability}} of the protein is critical for many applications, such as drug development. Predicting the stability of a protein can help determine whether the protein can perform its function efficiently in the cellular environment \cite{cheng2006prediction}.

\smallskip\noindent \textit{\textbf{Solubility}} reflects the solubility characteristics of a protein in a particular solvent. Predictions of solubility can help to understand whether a protein can exist and function properly within a cell \cite{hebditch2017protein}.

\smallskip\noindent \textit{\textbf{Mutation Effect Prediction}} of proteins refers to the assessment of the impact on various properties, structures, and functions of proteins when their amino acid sequences are changed \cite{mansoor2022accurate}. Commonly used methods include molecular dynamics-based methods, deep learning-based prediction models, and structural comparison methods.

\subsection{Sequence Generation Metrics} 
Protein sequence generation is the process of creating new protein sequences using specific methods, models, or algorithms \cite{anand2022protein}. Common evaluation methods include:

\smallskip\noindent \textit{\textbf{Perplexity (PPL)}} can be used to measure how accurately a model predicts amino acids \cite{hesslow2022rita}. The lower the perplexity, the more accurate the prediction.

\smallskip\noindent \textit{\textbf{Novelty}} refers to the degree of uniqueness of the generated protein sequence compared to a database of known protein sequences \cite{truong2023poet}.

\smallskip\noindent \textit{\textbf{Fr\'echet Protein Distance (FPD)}} is used to measure the similarity between the distribution represented by the generated protein sequence and the distribution of the real protein sequence~\citep{jiang2008protein}, denoted as: 
\begin{equation}
\delta_{\mathcal{F}}(f, g) = \inf_{\alpha, \beta} \max_{s \in [0,1]} \text{dist}(f(\alpha(s)), g(\beta(s)))
\end{equation}
where $\alpha$ and $\beta$ are continuous non-decreasing functions. The sequence distribution can be denoted by $f$ and $g$.

\smallskip\noindent \textit{\textbf{Diversity}} is designed to evaluate the degree of difference between a range of protein sequences generated by a model. Rich diversity means that the model is capable of generating a variety of different sequences. Common methods include Shannon Entropy \cite{bywater2015prediction} and Hamming Distance \cite{mcgee2021generative}.

\smallskip\noindent \textit{\textbf{Foldability}} focuses on whether the generated protein sequence can be folded into a stable three-dimensional structure. Measuring foldability is usually performed with tools such as RoseTTAFold \cite{baek2021accurate} or computational methods based on physicochemical principles \cite{magliery2015protein} to predict the likelihood that the generated sequence will form a stable structure.

\smallskip\noindent \textit{\textbf{Recovery}} is focused on the ability of a model to predict the corresponding sequence for a given structure accurately \cite{watson2023novo}. Evaluating recovery includes methods sequence comparison, structure comparison, functionality comparison, etc.

\end{document}